\pdfoutput=1
\documentclass[preprint,12pt]{elsarticle}

\usepackage[a4paper,margin=20mm]{geometry}
\usepackage{amsmath,amssymb}
\usepackage{bm}
\usepackage{graphicx}
\usepackage{xcolor}
\usepackage{float}
\usepackage{placeins}
\usepackage{hyperref}

\journal{Materials Today Quantum}

\begin{document}
\begin{frontmatter}

\title{Quantum magnetism of the spin-1 kagome-lattice antiferromagnet}

\author{Katsuhiro Morita}
\ead{katsuhiro.morita.c7@mso.tohoku.ac.jp}
\address{Department of Physics, Tohoku University, Sendai, Miyagi 980-8578, Japan}

\begin{abstract}
We investigate the spin-1 kagome-lattice Heisenberg antiferromagnet using large-scale Lanczos diagonalization and the finite-temperature Lanczos method. The zero-temperature magnetization process exhibits plateaus at $m=0$, $1/3$, $7/9$, and $8/9$, where $m$ is the normalized magnetization. The $m=0$ plateau is identified as a trimer valence-bond-crystal state, while the high-field plateaus at $m=7/9$ and $8/9$ are identified as magnon crystals. In particular, the $m=8/9$ plateau corresponds to the exact localized-magnon crystal state. A smoothed zero-temperature magnetization curve constructed using the Gaussian-kernel smoothing method indicates magnetization jumps at the lower-field edge of the $m=1/3$ plateau and at the upper-field edges of the $m=7/9$ and $8/9$ plateaus. At finite temperatures, the specific heat exhibits a double-peak structure with peaks around $T/J\simeq0.1$ and $T/J\simeq1.1$, and the low-temperature peak may be related to trimer valence-bond-crystal ordering. The finite-temperature magnetization curves show that the $m=1/3$ plateau remains visible at low temperatures, whereas the high-field plateaus are rapidly smeared out by thermal effects. These results provide benchmark data for thermodynamic and high-field magnetization measurements in candidate spin-1 kagome-lattice materials.
\end{abstract}


\end{frontmatter}

\section{Introduction}
The kagome-lattice quantum antiferromagnet provides a paradigmatic platform for exploring unconventional quantum states arising from geometric frustration~\cite{SpinL}. In this lattice, conventional magnetic ordering is strongly suppressed, and the competition among many nearly degenerate states can lead to nontrivial ground states~\cite{KL1,KL2}. In magnetic fields, frustrated quantum magnets can also exhibit magnetization plateaus and magnetization jumps, which are characteristic signatures of the interplay between frustration, quantum fluctuations, and magnetic field.

The spin-1/2 kagome-lattice antiferromagnet has been extensively studied theoretically. Using numerical methods such as exact diagonalization, density-matrix renormalization group (DMRG), infinite projected entangled-pair states (iPEPS), and variational Monte Carlo (VMC), its ground-state properties, magnetization process, and magnetization plateaus have been investigated in detail. Magnetization plateaus at normalized magnetizations $m = M/M_{\rm sat} = 0, 1/9, 1/3, 5/9,$ and $7/9$ in the magnetization process, where $M$ is the magnetization and $M_{\rm sat}$ is the saturation magnetization, have been discussed theoretically~\cite{KLMH1,KLMH2,KLMH3,KLMH4,KLMH5}. Various candidate states have been proposed for these plateaus, including $Z_2$ spin liquids~\cite{KLZ2-1,KLZ2-2,KLZ2-3}, $U(1)$ spin liquids~\cite{KLU1-1,KLU1-4,KLU1-5}, valence-bond crystals (VBCs)~\cite{KLVBC2,KLVBC3,KLMH8,1-9VBC1,1-9VBC2}, and magnon-crystal states~\cite{MCS1, KLMH7}. More recently, magnetization plateaus at $m=5/9$, $1/3$, and $1/9$ have also been reported experimentally in spin-1/2 kagome-lattice compounds~\cite{YCOHB1,YCOHB2,YCOHB3,YCOHB4,YCOHB6}, and the field-induced properties of kagome-lattice quantum magnets are now being actively studied from both theoretical and experimental perspectives.

In contrast, the spin-1 kagome-lattice antiferromagnet is expected to exhibit quantum states and field-induced phases that are distinct from those of the spin-1/2 system.
 For the zero-field ground state, several gapped nonmagnetic states have been discussed as candidates, including the hexagonal singlet solid~\cite{S1KHS1,S1KHS2} and the trimer VBC~\cite{KLMH5,S1KTVBC1,S1KTVBC2}. In magnetic fields, magnetization plateaus at $m = 1/3, 7/9,$ and $8/9$ have also been discussed~\cite{KLMH5,S1KMH}. In particular, the $m=8/9$ plateau is understood as an exact localized-magnon crystal state appearing just below the saturation magnetization. However, several issues remain unresolved, including the microscopic structures of the $m=1/3$ and $7/9$ plateaus, the possible presence of magnetization jumps, and thermodynamic responses at finite temperatures, such as the magnetic susceptibility, specific heat, and magnetization curves. Therefore, it is important to investigate the spin-1 kagome-lattice antiferromagnet using accurate large-scale numerical calculations.

Experimentally, interest in spin-1 kagome-lattice magnets has also been growing in recent years. Kagome-lattice and kagome-related compounds containing spin-1 ions, such as $\mathrm{Ni}^{2+}$ and $\mathrm{V}^{3+}$, have been synthesized and investigated~\cite{YCaVO,BaNiVO,VFluorides,YCaVO2,BondDisorder,NH4NiV,NaVF,BaNiVO2,NH4NiMo,BaNiAs}. 
Magnetic susceptibility and specific heat measurements have been used to characterize their thermodynamic properties, while high-field magnetization measurements have revealed field-induced behavior, including magnetization plateaus.
In fact, magnetization plateaus at $m = 1/3$ and $2/3$ have been reported in the $\mathrm{V}^{3+}$-based spin-1 kagome-lattice fluorides $\mathrm{Cs}_2\mathrm{KV}_3\mathrm{F}{12}$, $\mathrm{Cs}_2\mathrm{NaV}_3\mathrm{F}_{12}$, and $\mathrm{Rb}_2\mathrm{NaV}_3\mathrm{F}_{12}$~\cite{VFluorides}.
Plateau-like behavior around $m = 1/3$ has also been observed in $\mathrm{Ni}^{2+}$-based candidate spin-1 kagome compounds~\cite{BaNiVO2,BaNiAs}.

 In real materials, however, material-specific perturbations that are absent in the ideal nearest-neighbor Heisenberg model often play important roles. These include single-ion anisotropy, Dzyaloshinskii--Moriya interactions, further-neighbor interactions, and lattice distortions.
Therefore, to understand experimental results on candidate spin-1 kagome compounds, it is important to clarify the benchmark thermodynamic responses and high-field magnetization process of the ideal nearest-neighbor spin-1 kagome-lattice Heisenberg antiferromagnet. In particular, magnetic susceptibility, specific heat, and finite-temperature magnetization curves are experimentally accessible quantities and serve as important indicators for assessing whether real materials reflect ideal kagome Heisenberg physics or are strongly affected by material-specific perturbations. The finite-temperature and high-field theoretical data obtained in this study can thus serve as benchmark data for such comparisons.

In this study, we systematically investigate the zero-temperature and finite-temperature field-induced properties of the spin-1 kagome-lattice Heisenberg model using large-scale Lanczos diagonalization and the finite-temperature Lanczos method (FTLM)~\cite{Jaklic1994FTLM}. For the zero-temperature magnetization process, we determine the lowest energy in each magnetization sector for several finite clusters and examine the presence of magnetization plateaus and magnetization jumps. For finite-temperature properties, we calculate the magnetic susceptibility, specific heat, and finite-temperature magnetization curves using FTLM. Furthermore, to clarify the microscopic structures of the magnetization-plateau states, we analyze spin--spin correlations, spin structure factors, and dimer--dimer correlations.

\begin{figure}[tb]
\centering
\includegraphics[width=86mm]{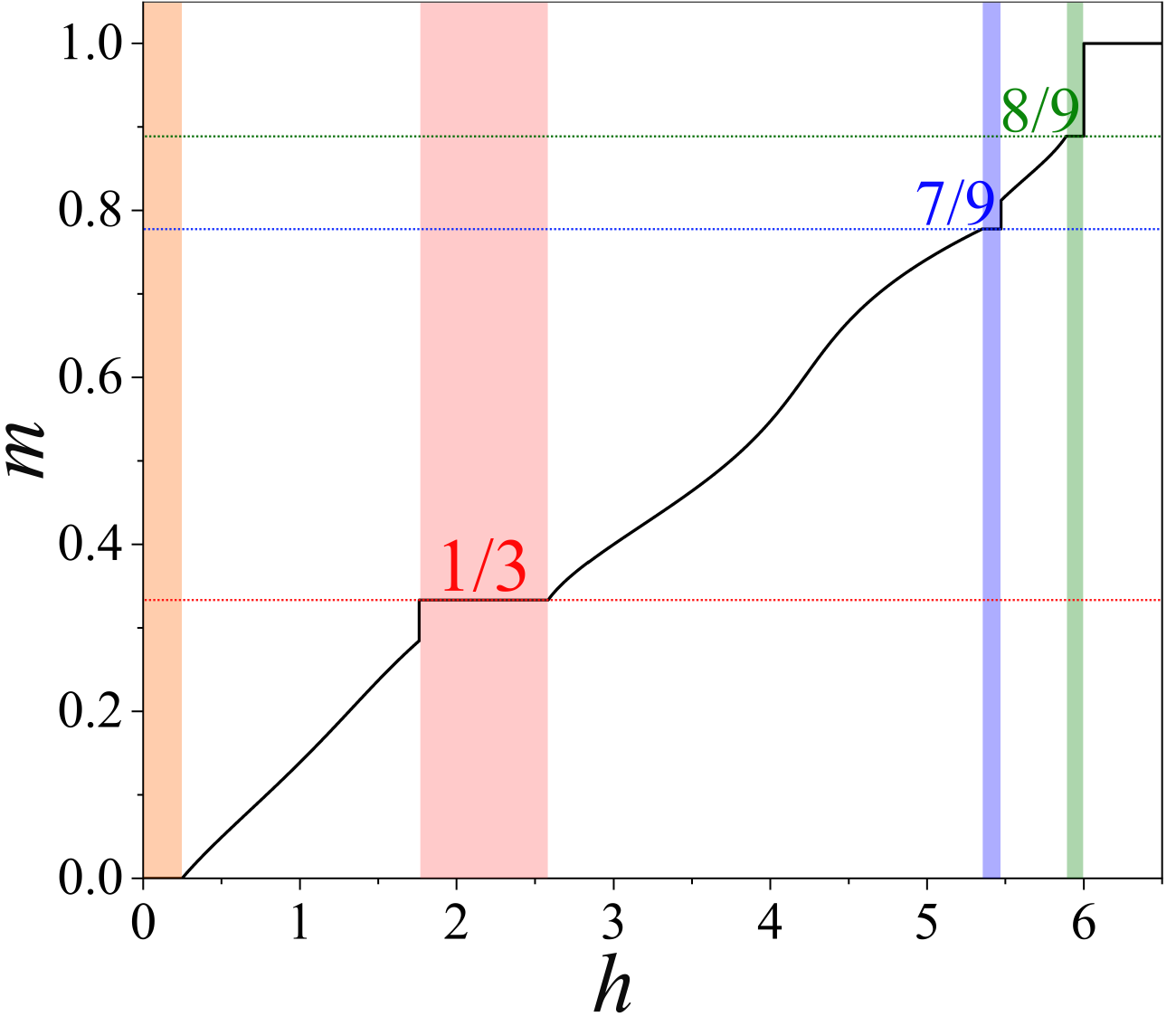}
\caption{
Zero-temperature magnetization curve of the spin-1 kagome-lattice Heisenberg antiferromagnet constructed using the Gaussian-kernel smoothing method from energy data obtained by Lanczos calculations.
 The colored regions indicate the magnetization plateaus at $m=0$, $1/3$, $7/9$, and $8/9$. 
Magnetization jumps appear on the lower-field side of the $m=1/3$ plateau and on the higher-field sides of the $m=7/9$ and $8/9$ plateaus.
}
\label{kerMH}
\end{figure}

Because the zero-temperature magnetization curves of finite-size systems have discrete staircase structures, we construct a smooth magnetization curve using a Gaussian-kernel smoothing method as an auxiliary tool to capture the overall magnetization process~\cite{GKm}. The resulting smoothed magnetization curve is shown in Fig.~\ref{kerMH}. Furthermore, by comparing this smoothed curve with the magnetization curves obtained by FTLM at low temperatures, we examine the validity of the smoothed magnetization curve and investigate the relationship between the zero-temperature magnetization process and the melting of magnetization plateaus at finite temperatures.

The main results obtained in this study are summarized as follows. As shown in Fig.~\ref{kerMH}, the zero-temperature magnetization process exhibits magnetization plateaus at $m = 0, 1/3, 7/9$, and $8/9$. Magnetization jumps are observed at the lower-field edge of the $m=1/3$ plateau, the upper-field edge of the $m=7/9$ plateau, and the upper-field edge of the $m=8/9$ plateau. At zero magnetization, the trimer VBC is identified; in particular, for the $N = 27$ cluster, an exact degeneracy associated with lattice-symmetry breaking is obtained. The high-field plateaus at $m = 7/9$ and $8/9$ can be understood as magnon-crystal states. Furthermore, using FTLM, we reveal a double-peak structure in the specific heat, the persistence of the $m=1/3$ plateau at low temperatures, and the melting of the plateaus with increasing temperature. 
These results provide benchmark data for interpreting magnetization plateaus, specific-heat anomalies, and field-induced phases in future studies of candidate spin-1 kagome-lattice compounds.

\section{Model and numerical methods}
\subsection{Model and zero-temperature Lanczos calculations}
We consider the spin-1 kagome-lattice Heisenberg model under an external magnetic field.
The Hamiltonian is given by
\begin{equation}
  \mathcal{H}
  = J \sum_{\langle i,j\rangle} \mathbf{S}_i \cdot \mathbf{S}_j
  - h \sum_i S_i^z ,
  \label{eq:hamiltonian}
\end{equation}
where $\mathbf{S}_i$ is the spin-1 operator at site $i$, $J>0$ is the nearest-neighbor antiferromagnetic exchange interaction, and $h$ is the external magnetic field. Hereafter, we set $J=1$ as the unit of energy.
 The total magnetization is defined as $M=\sum_i S_i^z$, and the saturation magnetization is $M_{\rm sat}=N$ for the spin-1 system, where $N$ is the number of sites. The normalized magnetization is then given by $m=M/M_{\rm sat}$.

The zero-temperature magnetization process is determined from the lowest energy $E_0(M)$ in each fixed-magnetization sector. In a magnetic field $h$, the energy of the sector with total magnetization $M$ is given by
\begin{equation}
E(M,h) = E_0(M) - hM .
\label{eq}
\end{equation}
Therefore, at each value of $h$, the ground state is obtained by minimizing $E(M,h)$ with respect to $M$, and the corresponding value of $M$ gives the zero-temperature magnetization.

We perform Lanczos calculations for several kagome-lattice clusters with periodic boundary conditions. To analyze the zero-temperature magnetization process, we use clusters with $N=24$, $27$, $30$, $36$, and $45$ sites. For clusters with $N \ge 30$, we mainly focus on the high-magnetization region. The geometries of the clusters are shown in the Supplementary Material.
For the large-scale parallel calculations, the MPI rank mapping was generated using the SCOTCH software package~\cite{SCOTCH}.

\subsection{Finite-temperature Lanczos calculations}

Finite-temperature properties are calculated using FTLM, which has been widely used for frustrated quantum  systems~\cite{FTLM1,FTLM2,FTLM3,FTLM4,FTLM5,FTLMM1,Ameltkagome1,FTLM6,FTLM7,FTLMM2,FTLM8,FTLMM3,FTLM9,
FTLM10,FTLM11}. In FTLM, the partition function and thermal averages are evaluated by stochastic trace sampling with random initial vectors. In the present model, the total magnetization $M$ is a conserved quantity. We therefore perform FTLM calculations separately in each fixed-magnetization sector.

Let $Z_M(T)$ denote the partition function of the zero-field Hamiltonian in the fixed-magnetization sector $M$. Since the magnetic field couples to the total magnetization as $-hM$, the total partition function and the thermal expectation value of the magnetization in a magnetic field are given by
\begin{equation}
  Z(T,h)  =  \sum_M e^{\beta h M} Z_M(T),
  \label{ZFTL}
\end{equation}
\begin{equation}
 M(T,h)  = \frac{\sum_M M e^{\beta hM} Z_M(T)}
{Z(T,h)} ,
\label{M}
\end{equation}
respectively, where $\beta=1/T$ and $k_{\rm B}=1$. The normalized magnetization is obtained as $m(T,h)=M(T,h)/M_{\rm sat}$. 
In the FTLM, the trace in each fixed-magnetization sector is approximated by stochastic sampling with random vectors, and the spectral decomposition is approximated in the Krylov subspace generated by the Lanczos procedure. The sector-resolved partition function is then written as
\begin{equation}
Z_M(T) \simeq
\frac{N_{\rm st}(M)}{R}
\sum_{r=1}^{R}
\sum_{j=0}^{N_{\rm L}-1}
e^{-\beta \epsilon_{j,M}^{(r)}}
\left|
\langle v_M^{(r)}|\psi_{j,M}^{(r)}\rangle
\right|^2 ,
\end{equation}
where $N_{\rm st}(M)$ is the dimension of the fixed-magnetization sector $M$, $R$ is the number of random vectors, $N_{\rm L}$ is the Krylov dimension, and $\epsilon_{j,M}^{(r)}$ and $|\psi_{j,M}^{(r)}\rangle$ are the Lanczos eigenvalues and eigenvectors obtained from the random vector $|v_M^{(r)}\rangle$.

In this study, we apply FTLM to clusters with $N=21$, $24$, and $27$ sites and calculate the magnetic susceptibility, specific heat, and finite-temperature magnetization curves. To improve the numerical accuracy, we use the replaced finite-temperature Lanczos method (RFTLM)~\cite{FTLMM1} for the $N=27$ cluster and the orthogonalized finite-temperature Lanczos method (OFTLM)~\cite{FTLMM1,FTLMM2,FTLMM3}  for the $N=21$ and $24$ clusters. Hereafter, RFTLM and OFTLM are collectively referred to as FTLM unless otherwise specified. The evaluation procedures for thermodynamic quantities, details of these improved schemes, and numerical conditions are described in the Supplementary Material.

\subsection{Correlation functions}

To investigate the microscopic structures of the magnetization-plateau states, we calculate nearest-neighbor spin--spin correlations, spin structure factors, and dimer--dimer correlations. The longitudinal spin structure factor is defined as
\begin{equation}
  S^{z}(\mathbf{q})
  =
  \frac{1}{N}
  \sum_{i,j}
  e^{i\mathbf{q}\cdot(\mathbf{r}_i-\mathbf{r}_j)}
  \left(
  \langle S_i^z S_j^z\rangle
  -
  \langle S_i^z\rangle \langle S_j^z\rangle
  \right),
  \label{eq:sqz}
\end{equation}
where $\mathbf{r}_i$ is the position vector of site $i$, and $\mathbf{q}$ is a wave vector allowed under the periodic boundary conditions of the kagome-lattice cluster.

In the kagome lattice, it is useful to distinguish between the reduced Brillouin zone associated with the three-site unit cell and the extended Brillouin zone constructed from the positions of all sites. In particular, the so-called $\mathbf{q}=\mathbf{0}$ order refers to an ordering pattern with wave vector $\mathbf{q}=\mathbf{0}$ in the Bravais lattice with a three-site unit cell. When the spin structure factor is represented in the extended Brillouin zone using all site positions, however, the corresponding spectral weight does not necessarily appear only at $\mathbf{q}=\mathbf{0}$.

The nearest-neighbor spin--spin correlation, hereafter referred to as the bond correlation, is defined as
\begin{equation}
  B_{ij}  =  \langle \mathbf{S}_i \cdot \mathbf{S}_j \rangle,
\end{equation}
where $i$ and $j$ are nearest-neighbor sites. We use this quantity to examine the trimer order in the zero-magnetization state.

To further investigate the spatial correlations of the bond correlations, we also calculate the dimer--dimer correlation function defined by
\begin{equation}
  D_{ij,kl}
  =
  \langle
  (\mathbf{S}_i \cdot \mathbf{S}_j)
  (\mathbf{S}_k \cdot \mathbf{S}_l)
  \rangle
  -
  B_{ij} B_{kl}.
  \label{eq:dimer_dimer}
\end{equation}
This quantity is used to identify the magnon-crystal state in the $m=7/9$ plateau.

\subsection{Magnetization curves using a Gaussian-kernel smoothing method}

The zero-temperature magnetization curves of finite-size systems exhibit discrete staircase structures because the total magnetization is quantized. In this study, as a tool to visualize the overall magnetization process, we approximate the energy density obtained in discrete magnetization sectors, $e(m) = \frac{E_0(M)}{N}$,
by a smooth function and construct a continuous $e$-$m$ curve. We then obtain a smoothed zero-temperature magnetization curve from this energy curve.

Specifically, when necessary, we divide the magnetization range into several regions and approximate the energy density in each region as
\begin{equation}
  e_{\rm ker}(m) = b_0 + b_1 m + b_2 m^2 + b_3 m^3
  + \sum_i a_i \left[ \exp\left[-\frac{(m-m_i)^2}{2\ell^2}\right] + \exp\left[-\frac{(m+m_i)^2}{2\ell^2}\right] \right].
  \label{kernel}
\end{equation}
Here, $m_i$ denotes the discrete normalized magnetization values corresponding to the fixed-magnetization sectors used in the Lanczos calculations, and $\ell$ is a hyperparameter controlling the kernel width. The coefficients $a_i$, $b_0$, $b_1$, $b_2$, and $b_3$ are determined so as to reproduce the energies in the fixed-magnetization sectors. For the obtained $e_{\rm ker}(m)$, we minimize
\begin{equation}
  e_{\rm ker}(m) - hm
  \label{eq:kernel_minimization}
\end{equation}
with respect to $m$ at each magnetic field $h$. The value of $m$ that gives the minimum is used to construct the smoothed zero-temperature magnetization curve.

This Gaussian-kernel smoothing method is used as a tool to make the overall magnetization process easier to grasp from the staircase-like magnetization curves obtained for finite-size systems~\cite{GKm}. The identification of magnetization plateaus and magnetization jumps is mainly based on the results obtained by the Lanczos calculations.
To examine the usefulness of this method, we also apply the same method to the spin-1/2 one-dimensional Heisenberg chain and the spin-1/2 triangular-lattice Heisenberg antiferromagnet. We confirm that the resulting magnetization curves agree well with the exact solution for the one-dimensional chain and with existing high-accuracy results for the triangular-lattice case. Details of the Gaussian-kernel smoothing method and these benchmark results are presented in the Supplementary Material.

\section{Results and discussion}

\subsection{Magnetization process}

\begin{figure}[tb]
\centering
\includegraphics[width=70mm]{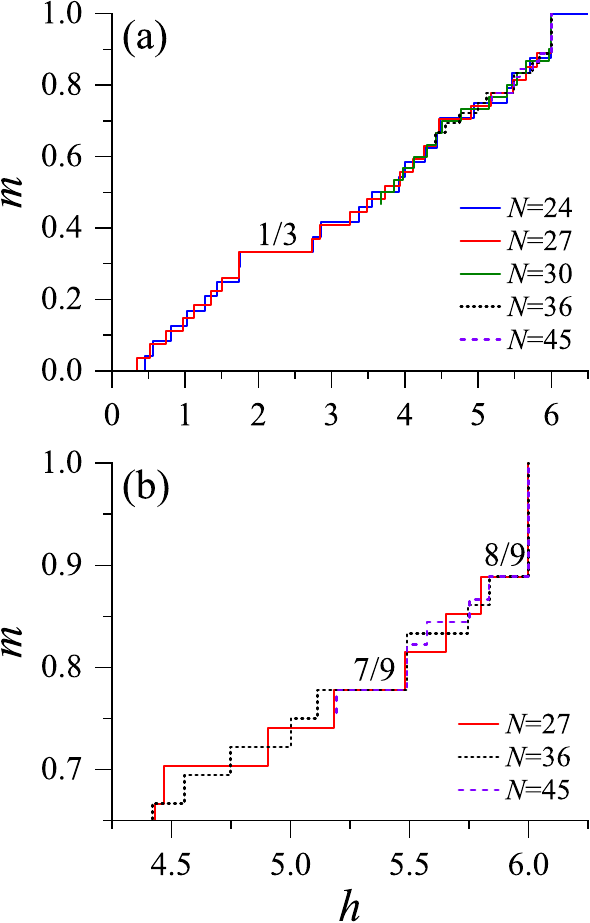}
\caption{
Zero-temperature magnetization process of the spin-1 kagome-lattice Heisenberg antiferromagnet obtained using the Lanczos method.
(a) Overall magnetization curves for the $N=24$, $27$, $30$, $36$, and $45$ clusters.
(b) Enlarged view of the high-magnetization region for the $N=27$, $36$, and $45$ clusters.
}
\label{EDMH}
\end{figure}

We first discuss the zero-temperature magnetization process. Figure~\ref{EDMH}(a) shows the zero-temperature magnetization curves of the spin-1 kagome-lattice Heisenberg antiferromagnet obtained by Lanczos calculations. In finite-size systems, the total magnetization $M$ takes discrete values, and therefore the magnetization curves exhibit staircase structures. In this study, we identify magnetization plateaus and magnetization jumps by combining the magnetization curves for several cluster sizes with the correlation-function analyses presented below.

Figure~\ref{EDMH}(a) shows that plateau structures appear at $m=0$ and $1/3$ in the low-field region. As discussed later based on the bond-correlation analysis, the $m=0$ plateau can be understood as the trimer VBC state.
In particular, at $m=1/3$, relatively wide flat regions are observed for both the $N=24$ and $27$ clusters, indicating the presence of a stable magnetization plateau. The structures near $m=7/9$ and $8/9$ in the high-magnetization region are less clearly resolved in the overall plot. Since $M_{\rm sat}=N$ for the spin-1 system, the system size must be a multiple of 9 in order to include the magnetization sectors corresponding exactly to $m=7/9$ and $8/9$. Therefore, Fig.~\ref{EDMH}(b) shows an enlarged view of the high-magnetization region for the $N=27$, $36$, and $45$ clusters.

The results for the $N=27$, $36$, and $45$ clusters shown in Fig.~\ref{EDMH}(b) indicate that the magnetization sectors corresponding to $m=7/9$ and $8/9$ become the ground state over finite magnetic-field ranges in the high-magnetization region. In particular, the $m=8/9$ plateau corresponds to the exact localized-magnon crystal state appearing just below the saturation magnetization. In strongly frustrated quantum spin systems, including the kagome lattice, spin flips from the fully polarized state can form localized magnons as exact eigenstates~\cite{7-9EX1,7-9EX2}. When these localized magnons are arranged without spatial overlap, a magnon-crystal state and a magnetization jump appear just below the saturation magnetization. Since the $m=8/9$ plateau in the spin-1 kagome lattice corresponds to such an exact localized-magnon crystal state, it can be understood as a stable high-field plateau even in the thermodynamic limit.
In contrast, the $m=7/9$ sector appears over a relatively wide magnetic-field range compared with the neighboring magnetization sectors, especially for the $N=36$ and $45$ clusters. 
Furthermore, the dimer--dimer correlations discussed later support the magnon-crystal character of this state. Therefore, the $m=7/9$ plateau can be understood as a magnon-crystal state analogous to the $m=8/9$ plateau, which corresponds to the exact localized-magnon crystal state.

Magnetization jumps can also be inferred from the behavior shown in Fig.~\ref{EDMH}. 
In Fig.~\ref{EDMH}(a), the magnetization increases steeply at the lower-field edge of the $m=1/3$ plateau for both the $N=24$ and $27$ clusters. We therefore regard this steep magnetization change as a finite-size precursor to first-order behavior in the thermodynamic limit. A related feature was reported in a previous iPEPS study; although the reported jump was not located exactly at the lower-field edge of the $m=1/3$ plateau, it appeared on the lower-field side of the plateau~\cite{KLMH5}. 
This consistency supports the interpretation that the steep magnetization change observed in the present Lanczos calculations reflects a first-order tendency on the lower-field side of the $m=1/3$ plateau.

In the high-magnetization region, as shown in Fig.~\ref{EDMH}(b), a tendency toward a magnetization jump is observed at the upper-field edge of the $m=7/9$ plateau. Although this behavior is not seen for the $N=27$ cluster, the larger $N=36$ and $45$ clusters show a skipped magnetization sector when the magnetization leaves the $m=7/9$ plateau. This tendency suggests that the magnetization jump at the upper-field edge of the $m=7/9$ plateau is not merely a small-cluster effect and may persist in the thermodynamic limit.
The magnetization jump at the upper-field edge of the $m=7/9$ plateau is reminiscent of the behavior at the upper-field edge of the $m=5/9$ plateau in the spin-1/2 kagome-lattice antiferromagnet. The $m=5/9$ plateau in the spin-1/2 system has been discussed as a magnon-crystal state. Although previous studies did not explicitly describe the upper-field-edge behavior of this plateau as a magnetization jump, both DMRG calculations with sine-square deformation and exact diagonalization results show a magnetization jump at this edge~\cite{KLMH3,KLMH4,KLMH6}.
 In the present spin-1 system, the dimer--dimer correlations discussed later also support the magnon-crystal  character of the $m=7/9$ plateau. These similarities suggest that the magnetization jump at the upper-field edge of the $m=7/9$ plateau may be associated with the destabilization of a magnon-crystal state, analogous to the spin-1/2 $m=5/9$ case.
For the $m=8/9$ plateau, the magnetization jump at its upper-field edge is a rigorous consequence of the exact localized-magnon crystal state and therefore persists in the thermodynamic limit. Thus, the high-magnetization process obtained in this study indicates that, in addition to the known $m=8/9$ localized-magnon crystal plateau, a magnetization jump may also occur at the upper-field edge of the $m=7/9$ plateau in the thermodynamic limit.

Finally, to visualize the overall magnetization process more clearly, we smooth the staircase-like magnetization curves obtained from finite-size Lanczos calculations using the Gaussian-kernel smoothing method. Figure~\ref{kerMH} shows the resulting smoothed zero-temperature magnetization curve. In this procedure, $m=0$, $1/3$, $7/9$, and $8/9$ are treated as candidate plateau positions. In contrast, the positions and presence of magnetization jumps are not assumed in advance. Instead, magnetization jumps appear as features obtained by minimizing $e_{\rm ker}(m)-hm$ with respect to $m$ at each magnetic field $h$, where $e_{\rm ker}(m)$ is the smoothed energy density obtained by the Gaussian-kernel smoothing method.
Thus, Fig.~\ref{kerMH} gives a clear overview of the zero-temperature magnetization process by showing a smoothed curve constructed from the finite-size Lanczos data, including the possible positions of magnetization jumps.
As shown in Fig.~\ref{kerMH}, the plateaus corresponding to $m=0$, $1/3$, $7/9$, and $8/9$ are retained in the smoothed magnetization curve. Magnetization jumps also appear near the lower-field edge of the $m=1/3$ plateau and the upper-field edge of the $m=7/9$ plateau. The reliability of this Gaussian-kernel smoothing method has been examined by comparing its results with the exact solution for the spin-1/2 one-dimensional Heisenberg chain and with existing high-accuracy results for the spin-1/2 triangular-lattice Heisenberg antiferromagnet. Details are presented in the Supplementary Material.
These results show that the zero-temperature magnetization process of the present model is characterized by plateau structures at $m=0$, $1/3$, $7/9$, and $8/9$, together with three magnetization jumps.

\subsection{Magnetic structures of plateau states}
In this subsection, we investigate the microscopic structures of the plateau states by analyzing bond correlations, spin structure factors, and dimer--dimer correlations.

\begin{figure}[tb]
\centering
\includegraphics[width=80mm]{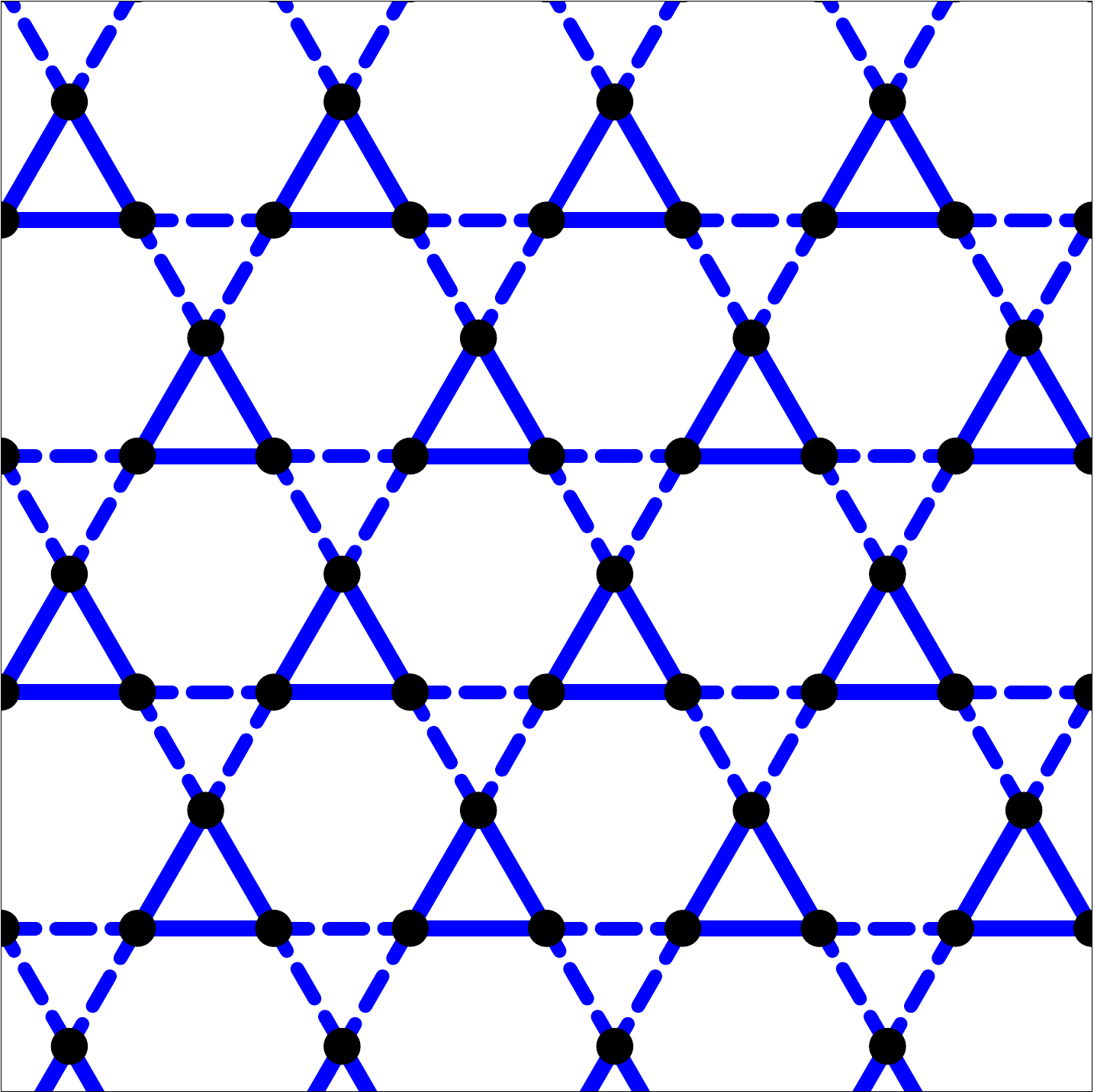}
\caption{
Bond correlations in the $m=0$ ground state for the $N=27$ cluster.
 To select one symmetry-broken component of the exactly twofold-degenerate ground-state subspace, 
the exchange interactions on the up triangles are increased from $J=1$ to $J=1+10^{-5}$. 
The solid and dashed lines represent bond correlations of $-0.7633$ and $-0.6552$, respectively.
}
\label{m0}
\end{figure}

\subsubsection{Trimer VBC at $m=0$}
We first discuss the zero-magnetization plateau state. For the zero-field ground state of the spin-1 kagome-lattice Heisenberg antiferromagnet, several gapped nonmagnetic states have been proposed as candidates, including the hexagonal singlet solid and the trimer VBC. In this study, we calculate bond correlations for the ground state obtained by the Lanczos method and analyze the microscopic structure of the zero-magnetization state.

For the $N=27$ cluster, we obtain an exact twofold degeneracy in the $m=0$ ground state. This degeneracy is associated with the rotational-symmetry breaking expected for the trimer VBC state. 
To visualize a single symmetry-broken component of this degenerate ground-state subspace, we introduce an infinitesimal symmetry-breaking perturbation only for the bond-correlation calculation shown in Fig.~\ref{m0}: the exchange interactions on the bonds forming the up triangles are increased from $J=1$ to $J=1+10^{-5}$.
Figure~\ref{m0} shows the resulting spatial distribution of the bond correlations in the $m=0$ state for the $N=27$ cluster. Strong antiferromagnetic bond correlations are concentrated on the up triangles of the kagome lattice, with bond-correlation values $-0.7633$ and $-0.6552$ for the strong and weak bonds, respectively. This result indicates that the bond correlations are not uniform over all bonds but form a trimerized pattern. This pattern preserves the translational symmetry of the lattice, while it distinguishes up and down triangles and reduces the sixfold rotational symmetry to threefold rotational symmetry. Thus, the $m=0$ state can be understood as the trimer VBC state with lattice-nematic character. To the best of our knowledge, the exact twofold degeneracy in the $N=27$ periodic cluster has not been explicitly demonstrated in previous numerical studies.
These results indicate that the zero-magnetization plateau in the present model is not merely due to a finite-size gap, but can be understood as a nonmagnetic state accompanied by trimer order. In particular, the exact twofold degeneracy and the spatial pattern of the bond correlations for the $N=27$ cluster strongly support the possibility that the trimer VBC state remains stable in the thermodynamic limit.

\subsubsection{Magnon-crystal states at $m=7/9$ and $8/9$}

\begin{figure}[tb]
\centering
\includegraphics[width=120mm]{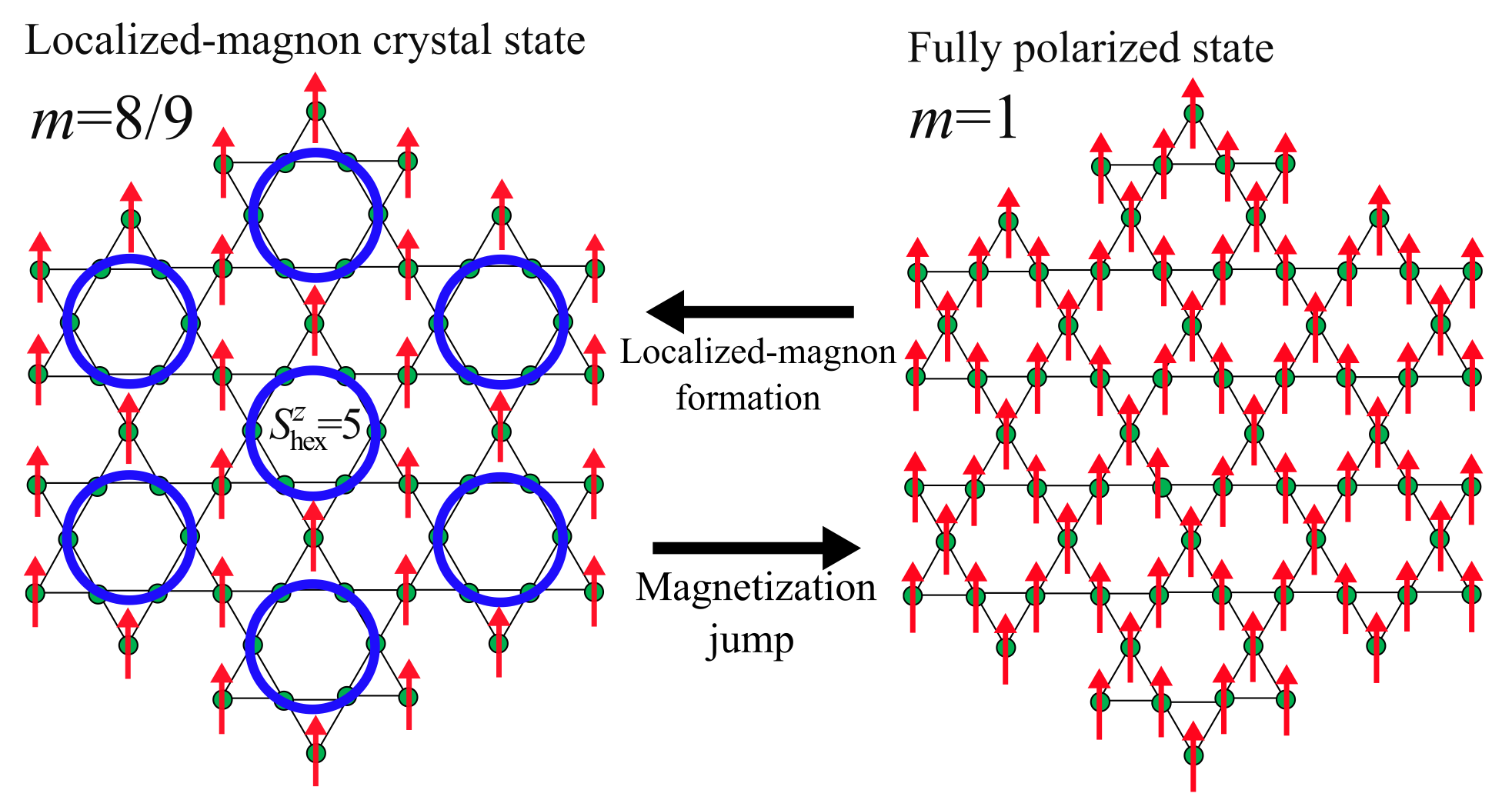}
\caption{
Schematic illustration of the localized-magnon crystal state at $m=8/9$ and the fully polarized state at $m=1$. Each blue circle represents a localized magnon on a hexagon with $S_{\rm hex}^z=5$. 
The transition from the $m=8/9$ plateau to the fully polarized state is accompanied by a magnetization jump.
}
\label{LMC}
\end{figure}

We next discuss the magnetic structures of the $m=7/9$ and $8/9$ plateaus appearing in the high-magnetization region. First, the $m=8/9$ plateau is understood as the exact localized-magnon crystal state. Indeed, in the magnetization process obtained by the present Lanczos calculations, a magnetization jump to the saturation magnetization is observed at the upper-field edge of the $m=8/9$ plateau. Since the wave function of this state is already known, we do not perform additional correlation-function calculations for the $m=8/9$ state in this study.
Figure~\ref{LMC} schematically illustrates the formation of localized magnons from the fully polarized state and the resulting magnon-crystal state.

\begin{figure}[tb]
\centering
\includegraphics[width=80mm]{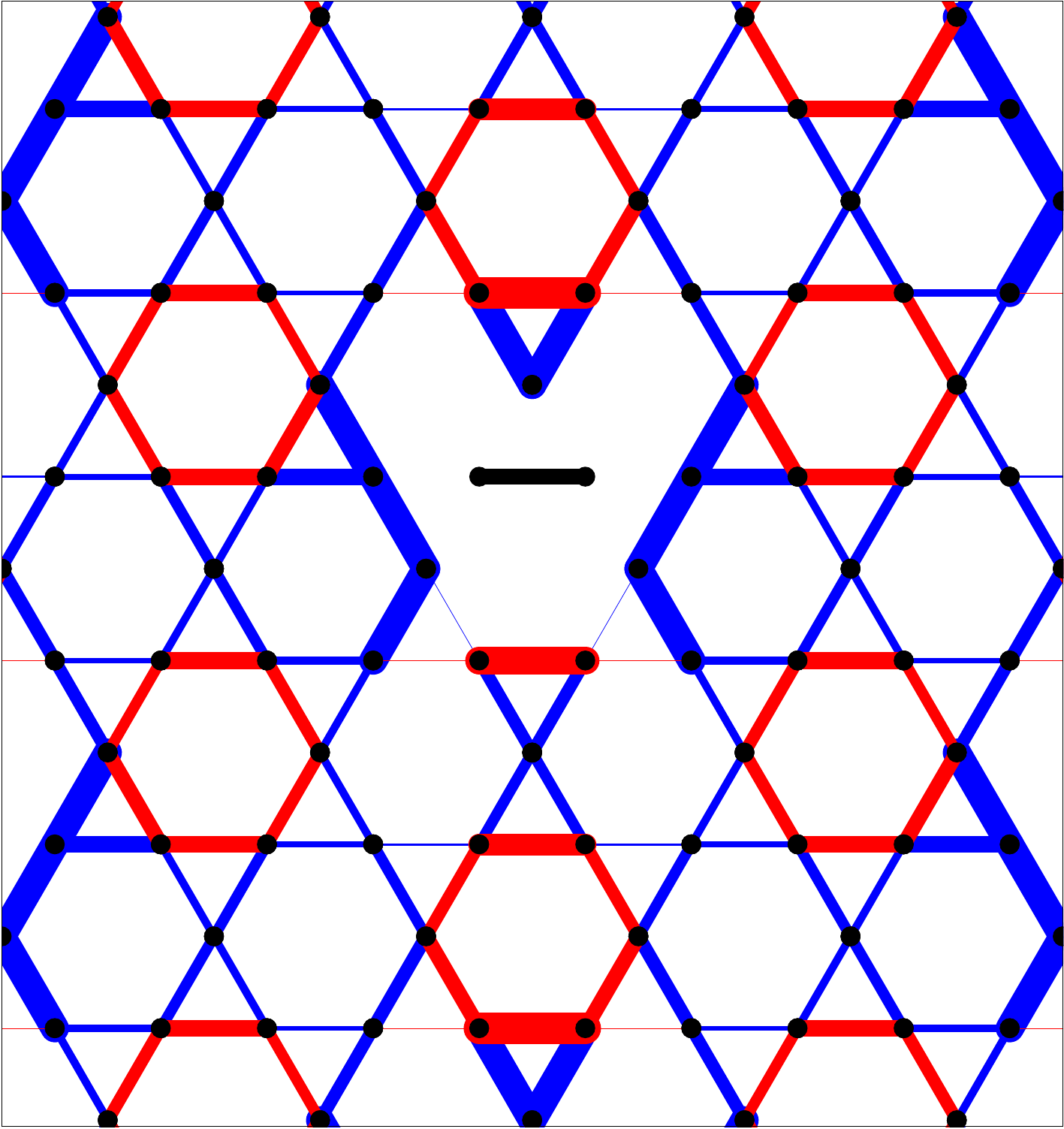}
\caption{
Dimer--dimer correlations in the $m=7/9$ ground state for the $N=36$ cluster. The black line represents the reference dimer. Red and blue lines indicate positive and negative correlations, respectively. The line thickness represents the magnitude of the correlation. Hexagons with positive correlations are periodically arranged, indicating that the $m=7/9$ plateau has magnon-crystal character.
}
\label{MCS}
\end{figure}

A key finding of this study is that the $m=7/9$ plateau can also be understood as a magnon-crystal state. Figure~\ref{MCS} shows the dimer--dimer correlations of the $m=7/9$ state obtained for the $N=36$ cluster. The correlations exhibit a periodic pattern in which hexagons with positive correlations are regularly arranged. This periodic correlation pattern is distinct from that expected for a spatially uniform high-magnetization state, and indicates that magnons are strongly localized on specific hexagons and form a regular spatial arrangement. Therefore, although the $m=7/9$ plateau is different from the exact localized-magnon crystal state at $m=8/9$, it can also be interpreted as a magnon-crystal state.
This interpretation is also consistent with the magnetization jumps observed in the high-magnetization region. The $m=8/9$ plateau is the exact localized-magnon crystal state, and the magnetization jump to the saturation magnetization at its upper-field edge follows rigorously from this exact state. The dimer--dimer correlations obtained in this study indicate that the $m=7/9$ plateau also exhibits magnon-crystal order similar to that of the $m=8/9$ plateau. Therefore, it is natural to interpret the magnetization jump at the upper-field edge of the $m=7/9$ plateau as arising from the destabilization of this magnon-crystal order and the resulting first-order transition to another magnetization state.
This magnon-crystal structure of the $m=7/9$ plateau closely resembles the structure reported for high-magnetization plateaus in the spin-1/2 kagome-lattice antiferromagnet. In particular, for the $m=5/9$ plateau in the spin-1/2 system, a periodic arrangement of hexagons with strong correlations has been reported in the dimer--dimer correlations~\cite{KLMH3}. The $m=7/9$ plateau in the present spin-1 system exhibits a similar spatial pattern. Therefore, the $m=7/9$ plateau can be understood as a magnon-crystal state analogous to the $m=5/9$ plateau in the spin-1/2 system.
These results indicate that the high-magnetization plateaus at $m=7/9$ and $8/9$ can be understood in a unified manner as magnon-crystal states. They are field-induced phases stabilized by a mechanism distinct from that of the trimer VBC state in the low-field region.

\subsubsection{Magnetic structure of the $m=1/3$ plateau}

Finally, we discuss possible magnetic structures of the $m=1/3$ plateau. 
As shown in Figs.~\ref{kerMH} and \ref{EDMH}(a), the $m=1/3$ plateau is stabilized over a relatively wide magnetic-field range in the zero-temperature magnetization process. It is therefore one of the prominent magnetization plateaus of the present model. In frustrated magnets based on triangular units, the $m=1/3$ plateau is often associated with an up-up-down (uud) magnetic structure. For the kagome lattice, natural candidates include the $\mathbf{q}=\mathbf{0}$ uud structure preserving the three-site unit cell, the $\sqrt{3}\times\sqrt{3}$ uud structure with a larger periodicity, and a magnon-crystal-like state similar to that found in the $m=7/9$ plateau.

\begin{figure}[tb]
\centering
\includegraphics[width=70mm]{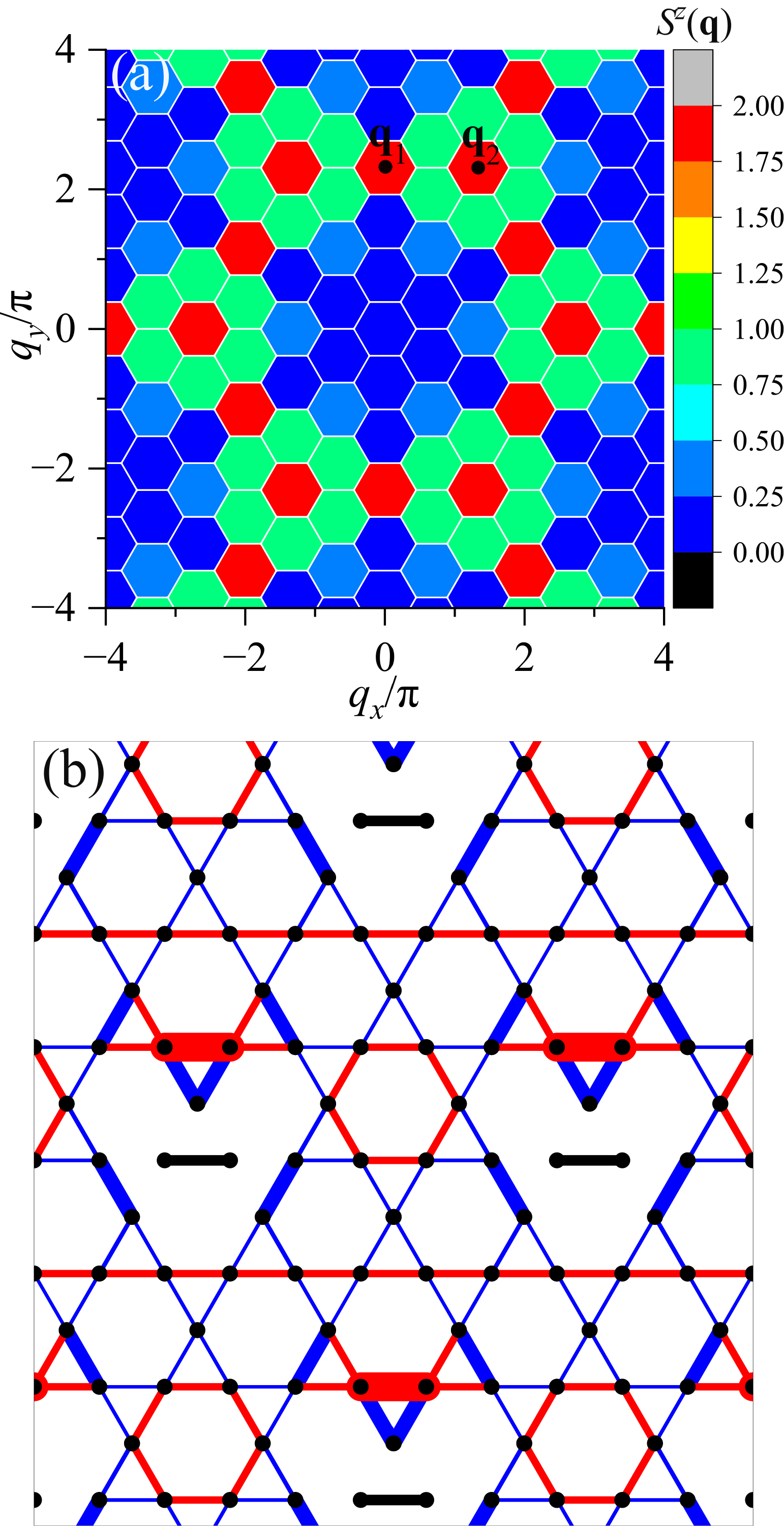}
\caption{
Magnetic correlations in the $m=1/3$ ground state for the $N=27$ cluster.
(a) Longitudinal spin structure factor $S^z(\mathbf{q})$.
The labels $\mathbf{q}_1$ and $\mathbf{q}_2$ denote representative wave-vector positions corresponding to the $\mathbf{q}=\mathbf{0}$ uud structure and the $\sqrt{3}\times\sqrt{3}$ uud structure, respectively.
The corresponding intensities are $S^z(\mathbf{q}_1)=1.996$ and $S^z(\mathbf{q}_2)=1.870$, indicating that the intensity at $\mathbf{q}_1$ is slightly larger than that at $\mathbf{q}_2$.
(b) Dimer--dimer correlations. The black line represents the reference dimer. Red and blue lines indicate positive and negative correlations, respectively. The line thickness represents the magnitude of the correlation.
}
\label{13Sq}
\end{figure}

We therefore calculate the spin structure factor and dimer--dimer correlations for the $m=1/3$ plateau state. Figure~\ref{13Sq}(a) shows the spin structure factor in the $m=1/3$ state for the $N=27$ cluster, and Fig.~\ref{13Sq}(b) shows the corresponding dimer--dimer correlations. The spin structure factor shows enhanced intensities at the wave-vector positions labeled $\mathbf{q}_1$ and $\mathbf{q}_2$ in Fig.~\ref{13Sq}(a). These positions correspond to the characteristic peak positions expected for the $\mathbf{q}=\mathbf{0}$ uud structure and the $\sqrt{3}\times\sqrt{3}$ uud structure, respectively. The intensity at $\mathbf{q}_1$ is slightly larger than that at $\mathbf{q}_2$. 
The dimer--dimer correlations, however, show a different tendency: positive correlations are enhanced on some hexagons, suggesting possible magnon-crystal character. In the $N=27$ cluster, among the hexagons that do not include the reference dimer, there are two hexagons where strong positive correlations are expected for an ideal magnon-crystal state. In the present calculation, however, strong positive correlations appear only on one of them. Thus, this pattern alone is not sufficient to conclude that a fully developed magnon-crystal structure is realized.

The spin structure factor shows comparable intensities at the wave-vector positions associated with the $\mathbf{q}=\mathbf{0}$ and $\sqrt{3}\times\sqrt{3}$ uud structures. The dimer--dimer correlations also show only a partial magnon-crystal-like pattern. Therefore, the microscopic structure of the $m=1/3$ plateau cannot be uniquely determined from the present finite-size data.
A similar ambiguity has also been reported in previous studies. For example, iPEPS studies have proposed both the $\mathbf{q}=\mathbf{0}$ uud structure and the $\sqrt{3}\times\sqrt{3}$ uud structure as candidates for the $m=1/3$ plateau~\cite{KLMH5}. These results indicate that a complete consensus on the magnetic structure of this plateau has not yet been reached. Further numerical studies for larger systems are required to clarify this issue.

\subsection{Thermodynamic properties}
We next discuss the thermodynamic properties obtained using FTLM.
Systematic calculations of the magnetic susceptibility, specific heat, and finite-temperature magnetization curves for the spin-1 kagome-lattice Heisenberg antiferromagnet are important for future comparisons with candidate spin-1 kagome-lattice materials. Since these quantities are routinely measured in many magnetic compounds, the finite-temperature data obtained in this study can serve as useful benchmark data for interpreting experimental results.

\begin{figure}[tb]
\centering
\includegraphics[width=160mm]{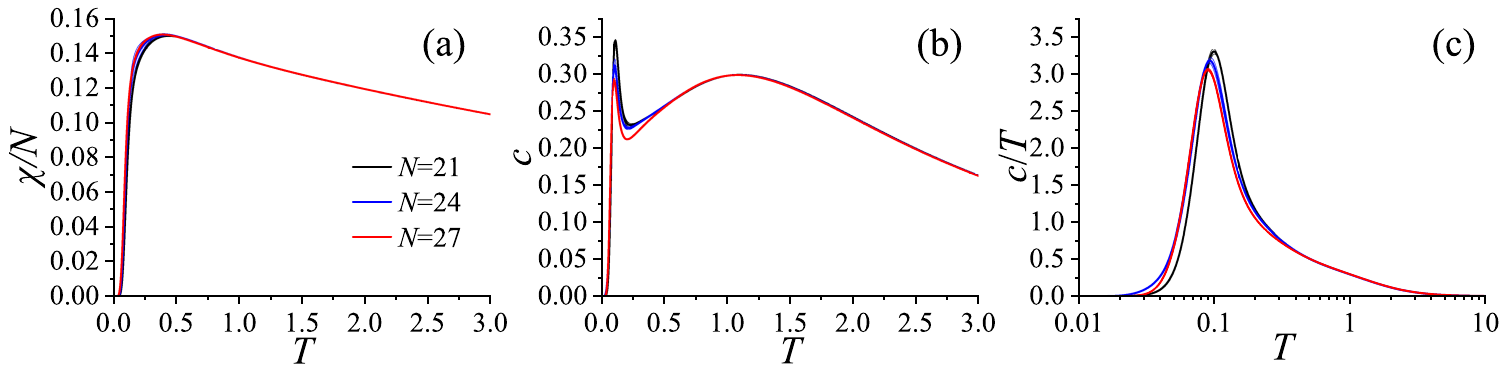}
\caption{
Zero-field thermodynamic quantities for the $N=21$, $24$, and $27$ clusters obtained using FTLM. (a) Magnetic susceptibility per site, $\chi/N$. (b) Specific heat per site, $c=C/N$. (c) Specific heat divided by temperature, $c/T$. The shaded regions indicate the standard errors estimated from the random-vector sampling; in most cases, they are smaller than the line width.
}
\label{FTQ}
\end{figure}

\subsubsection{Magnetic susceptibility and specific heat}
Figure~\ref{FTQ}(a) shows the magnetic susceptibility per site, $\chi/N$, obtained using FTLM. The calculations were performed for the $N=21$, $24$, and $27$ clusters. In the high-temperature region, the susceptibility increases as the temperature is lowered. Upon further lowering the temperature, it exhibits a broad maximum around $T\simeq 0.4$ and then rapidly decreases in the low-temperature region. For $T\gtrsim 0.4$, the results for $N=21$, $24$, and $27$ almost overlap, indicating that the size dependence is small in this temperature range. In addition, the results for the $N=24$ and $27$ clusters agree well over almost the entire temperature range. These results suggest that, at least for the magnetic susceptibility, finite-size effects are relatively small within the cluster sizes considered here. The calculated susceptibility therefore reasonably reflects the thermodynamic-limit behavior.

Figure~\ref{FTQ}(b) shows the specific heat per site, $c=C/N$. The most characteristic feature is the appearance of a double-peak structure for all cluster sizes, $N=21$, $24$, and $27$. A sharp low-temperature peak appears around $T\simeq 0.1$, while a broad high-temperature peak is formed around $T\simeq 1.1$. 
The high-temperature peak in the specific heat and the broad maximum in the magnetic susceptibility are both related to the development of short-range antiferromagnetic correlations, although they reflect different physical quantities. The high-temperature peak in the specific heat indicates the temperature region where the internal energy shows a relatively large temperature dependence. In contrast, the broad maximum in the magnetic susceptibility arises because antiferromagnetic correlations become sufficiently developed upon cooling and suppress the magnetic response. 

The sharp low-temperature peak may reflect the formation of the trimer VBC state.
For the specific heat, the results for $N=21$, $24$, and $27$ also almost overlap for $T\gtrsim 0.4$, indicating small size dependence in this temperature range. In contrast, the height of the low-temperature peak shows a clearer size dependence, although the peak position is relatively stable. This suggests that the low-temperature energy scale itself is robust, while the peak height and width may still change quantitatively toward the thermodynamic limit.

To examine the low-temperature structure more clearly, Fig.~\ref{FTQ}(c) shows $c/T$. The temperature axis is plotted on a logarithmic scale to highlight the low-temperature region. 
Since $c/T$ corresponds to the temperature derivative of the entropy per site, $s=S/N$, its peak indicates the temperature range where entropy is released most intensively. As shown in Fig.~\ref{FTQ}(c), a pronounced peak appears around $T\simeq 0.1$. This peak occurs in almost the same temperature region for all cluster sizes, $N=21$, $24$, and $27$, indicating the presence of a well-defined low-temperature energy scale associated with entropy release.
Furthermore, the size dependence of $c/T$ is small around $T\simeq 0.4$, suggesting that the values in this temperature region reasonably reflect the thermodynamic-limit behavior. As the temperature is lowered from this region, $c/T$ increases toward the low-temperature peak. Since the zero-magnetization ground state is expected to be a gapped nonmagnetic state, $c/T$ should eventually decrease again and approach zero at sufficiently low temperatures. Therefore, the present results suggest that a peak in $c/T$ remains in the low-temperature region below $T\simeq 0.4$ even in the thermodynamic limit.
These results indicate that the low-temperature structure of the specific heat is not merely a finite-size artifact, but is likely to remain in the thermodynamic limit. Thus, the specific heat of the spin-1 kagome-lattice Heisenberg antiferromagnet is expected to exhibit a double-peak structure.
The low-temperature peak in the specific heat is also interesting in relation to the trimer VBC state. As discussed in the previous subsection, the ground state has a trimer VBC structure that breaks the rotational symmetry of the lattice. An ordered state associated with such a discrete lattice-symmetry breaking can, in principle, undergo a finite-temperature phase transition in two dimensions. Indeed, a finite-temperature phase transition has been discussed for the plaquette VBC phase in the Shastry--Sutherland model, which also involves lattice-symmetry breaking~\cite{SSSB1,SSSB2}. Therefore, the low-temperature peak observed in $c$ in the present study may reflect a finite-temperature transition associated with the formation of trimer VBC order, rather than a simple crossover due only to low-energy excitations.
However, the present FTLM calculations are based on finite clusters, and it is difficult to determine the existence of a true phase transition solely from the sharpness and size dependence of the peaks. Larger-scale finite-temperature calculations are required to establish whether such a transition occurs in the thermodynamic limit.

\begin{figure}[tb]
\centering
\includegraphics[width=70mm]{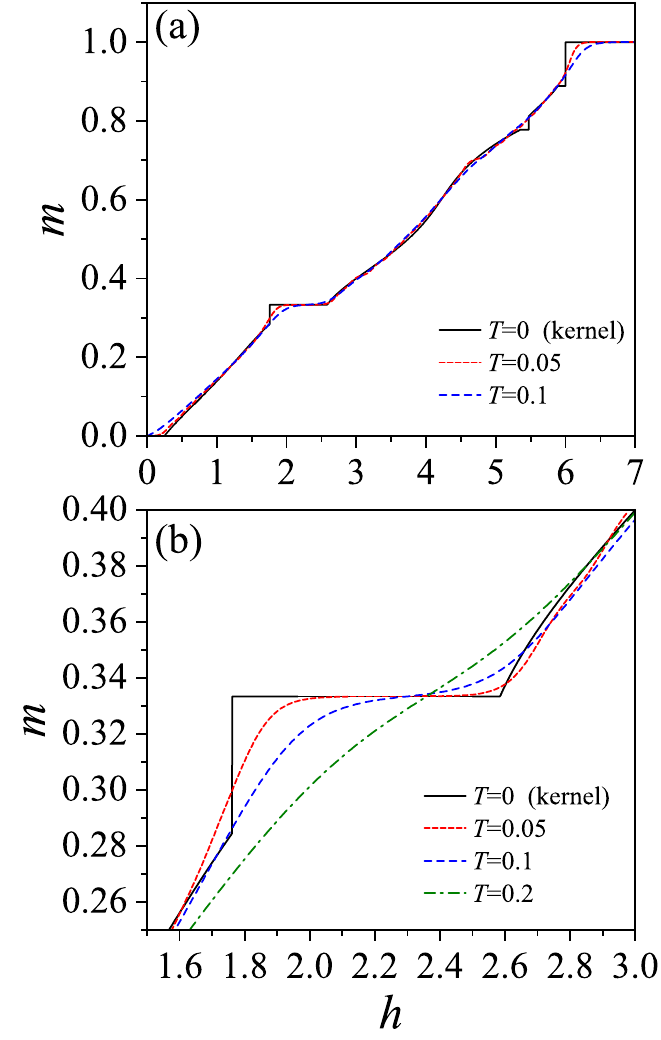}
\caption{
Finite-temperature magnetization curves. 
(a) Comparison between the low-temperature magnetization curves for the $N=27$ cluster obtained using FTLM and the zero-temperature magnetization curve obtained using the Gaussian-kernel smoothing method. 
(b) Enlarged view around the $m=1/3$ plateau.
}
\label{FTMH}
\end{figure}

\subsubsection{Finite-temperature magnetization curves}
We next discuss the finite-temperature magnetization curves. By examining the magnetization curves at finite temperatures, we can clarify how the magnetization plateaus appearing at zero temperature are thermally smeared out as the temperature increases.

Figure~\ref{FTMH}(a) compares the low-temperature magnetization curves for the $N=27$ cluster obtained using FTLM with the zero-temperature magnetization curve obtained by the Gaussian-kernel smoothing method. The size dependence of the finite-temperature magnetization curves is shown in the Supplementary Material. At $T=0.1$, the results for $N=21$, $24$, and $27$ almost coincide, indicating relatively small finite-size effects. At $T=0.05$, however, residual finite-size effects remain in the low-temperature plateau structures and in the high-magnetization region. Thus, quantitative discussion at this temperature should be made with some care.

The magnetization curves at $T=0.05$ and $T=0.1$ broadly follow the zero-temperature magnetization process. In particular, the structure corresponding to the $m=1/3$ plateau remains visible even at finite temperatures, showing that the main plateau structure obtained at zero temperature is reflected in the finite-temperature magnetization curves.
In contrast, the high-magnetization plateaus at $m=7/9$ and $8/9$ do not show clear flat regions in the finite-temperature curves, even at $T=0.05$. This is probably because these plateaus have relatively narrow widths and are easily smeared out by thermal effects. 
Since calculations at still lower temperatures suffer from pronounced finite-size staircase structures originating from the zero-temperature magnetization process, we do not discuss the finite-temperature melting of these high-magnetization plateaus quantitatively.
Nevertheless, the present results suggest that temperatures below $T/J=0.05$ would be required to observe these plateaus clearly in experiments.

Figure~\ref{FTMH}(b) shows an enlarged view around the $m=1/3$ plateau. The result for $T=0.2$ is also shown to examine how the plateau is thermally smeared out. At $T=0.05$, a flat region close to the zero-temperature plateau remains visible. At $T=0.1$, the plateau edges become clearly rounded. At $T=0.2$, a distinct flat region is almost lost. These results indicate that temperatures below $T/J \simeq 0.1$ are required to identify the $m=1/3$ plateau in experimental magnetization curves. For a more precise determination of the plateau width and its edges, measurements at lower temperatures, $T/J\lesssim 0.05$, are desirable.

The thermal smearing of the $m=1/3$ plateau is not completely symmetric. As shown in Fig.~\ref{FTMH}(b), the rounding behavior differs slightly between the lower- and upper-field sides of the plateau. In the zero-temperature magnetization process, the magnetization jump appears near the lower-field edge of the $m=1/3$ plateau. This low-field-edge structure may affect the asymmetric thermal smearing of the plateau at finite temperatures.
A similar asymmetric thermal smearing has also been discussed for the spin-1/2 kagome-lattice antiferromagnet. In the present spin-1 system, however, the asymmetry in the finite-temperature magnetization curves is relatively weak, and its quantitative behavior differs from that in the spin-1/2 system. This difference suggests that the stability of the $m=1/3$ plateau, the surrounding low-energy excitations, and the microscopic structure of the plateau state depend on the spin quantum number. 

These results show that finite-temperature magnetization curves provide a useful way to examine how the zero-temperature plateau structures survive and are thermally smeared out at finite temperatures. In particular, the thermal smearing of the $m=1/3$ plateau and its weak asymmetry are characteristic features that can be tested experimentally.

\section{Conclusions}

In this study, we systematically investigated the zero-temperature magnetization process, zero-field thermodynamic properties, and finite-temperature magnetization curves of the spin-1 kagome-lattice Heisenberg antiferromagnet using large-scale Lanczos diagonalization and FTLM. From the lowest energies in fixed-magnetization sectors, finite-size magnetization curves, and correlation-function analyses, we identified magnetization plateaus at $m=0$, $1/3$, $7/9$, and $8/9$. The $m=0$ and $1/3$ plateaus are clearly observed in the low-field magnetization process, whereas the $m=7/9$ plateau is supported by both the high-field magnetization behavior and the dimer--dimer correlations. The $m=8/9$ plateau corresponds to the known exact localized-magnon crystal state. We also used the Gaussian-kernel smoothing method to obtain the smoothed zero-temperature magnetization curve. This analysis reveals magnetization jumps not only at the upper-field edge of the known $m=8/9$ plateau, but also at the lower-field edge of the $m=1/3$ plateau and at the upper-field edge of the $m=7/9$ plateau. The resulting zero-temperature curve is consistent with the low-temperature FTLM magnetization curves.

We also clarified the microscopic structures of the plateau states. At zero magnetization, several nonmagnetic states have been proposed as candidates, including the hexagonal singlet solid and the trimer VBC state. Our bond-correlation analysis identifies the $m=0$ plateau as the trimer VBC state, because the bond correlations exhibit a trimerized spatial pattern that breaks the rotational symmetry of the lattice. For the $N=27$ cluster, we found an exact twofold degeneracy associated with this symmetry breaking, providing strong support for the trimer VBC state. In the high-magnetization region, the $m=7/9$ and $8/9$ plateaus can be understood as field-induced magnon-crystal states. In particular, the dimer--dimer correlations in the $m=7/9$ plateau exhibit a periodic arrangement of hexagons with strong positive correlations, indicating magnon-crystal character.

For the $m=1/3$ plateau, we analyzed the spin structure factor and dimer--dimer correlations. In the finite clusters studied here, the spin structure factor shows comparable enhanced intensities at the wave-vector positions associated with the $\mathbf{q}=\mathbf{0}$ and $\sqrt{3}\times\sqrt{3}$ uud structures, while the dimer--dimer correlations show only a partial magnon-crystal-like pattern. These results do not allow us to uniquely identify the microscopic structure of the $m=1/3$ plateau. This ambiguity is consistent with previous numerical studies, where different candidate structures have been proposed for the $m=1/3$ plateau. Further calculations for larger systems are required to clarify the microscopic structure of this plateau.

We further studied finite-temperature properties for the $N=21$, $24$, and $27$ clusters using FTLM. The magnetic susceptibility exhibits a broad maximum around $T\simeq 0.4$ and then decreases rapidly at low temperatures. The results show small size dependence for $T\gtrsim 0.4$, and the calculated susceptibility appears to reflect the thermodynamic-limit behavior well. The specific heat exhibits a double-peak structure for all cluster sizes, with a low-temperature peak around $T/J\simeq 0.1$ and a broad high-temperature peak around $T/J\simeq 1.1$. 
The high-temperature peak in the specific heat and the broad maximum in the magnetic susceptibility are both related to the development of short-range antiferromagnetic correlations, although they reflect different physical quantities. The high-temperature peak in the specific heat indicates the temperature region where the internal energy shows a relatively large temperature dependence. In contrast, the broad maximum in the magnetic susceptibility arises because antiferromagnetic correlations become sufficiently developed upon cooling and suppress the magnetic response. 
The low-temperature peak in the specific heat may reflect the formation of trimer VBC order and may be related to a finite-temperature transition. Larger-scale calculations are necessary to establish whether a true phase transition occurs in the thermodynamic limit.

Finally, we calculated finite-temperature magnetization curves. The $m=1/3$ plateau remains visible at low temperatures, while its plateau edges are thermally rounded as the temperature increases. The thermal smearing of this plateau is slightly asymmetric between the lower- and upper-field sides, which may be related to the magnetization jump at the lower-field edge in the zero-temperature magnetization process. The high-magnetization plateaus at $m=7/9$ and $8/9$ are more easily smeared out by thermal effects because of their narrow plateau widths. The present results suggest that temperatures below $T/J=0.05$ would be required to observe these high-magnetization plateaus clearly in experiments.

The present results provide important benchmark data for future thermodynamic and high-field magnetization measurements on candidate spin-1 kagome-lattice materials. They are expected to be useful for interpreting experimental results and for identifying spin-1 kagome-lattice antiferromagnetic materials.

\section*{Acknowledgements}
The author thanks T. Nakamura and H. Ueda for valuable discussions. 
This work was supported by JSPS KAKENHI Grant Number JP25K07230. The numerical calculations were performed using the facilities of the Supercomputer Center, the Institute for Solid State Physics, The University of Tokyo, and the supercomputer Fugaku provided by RIKEN through the HPCI System Research Project (Project ID: hp250573).

\section*{Declaration of generative AI and AI-assisted technologies in the writing process}
During the preparation of this work, the author used ChatGPT (OpenAI) to assist with language editing and readability improvement. After using this tool, the author reviewed and edited the content as needed and takes full responsibility for the content of the published article.

\bibliographystyle{elsarticle-num}
\bibliography{refK}

\clearpage
\suppressfloats[t]

\setcounter{figure}{0}
\renewcommand{\thefigure}{S\arabic{figure}}

\setcounter{table}{0}
\renewcommand{\thetable}{S\arabic{table}}

\setcounter{equation}{0}
\renewcommand{\theequation}{S\arabic{equation}}

\begin{center}
\Large
Supplementary Material for\\[2mm]
Quantum magnetism of the spin-1 kagome-lattice antiferromagnet

\vspace{6mm}

\normalsize
Katsuhiro Morita
\\
\vspace{2mm}

\small
Department of Physics, Tohoku University, Sendai, Miyagi 980-8578, Japan
\end{center}

\section{Cluster geometries}
\begin{figure}[H]
\centering
\includegraphics[width=110mm]{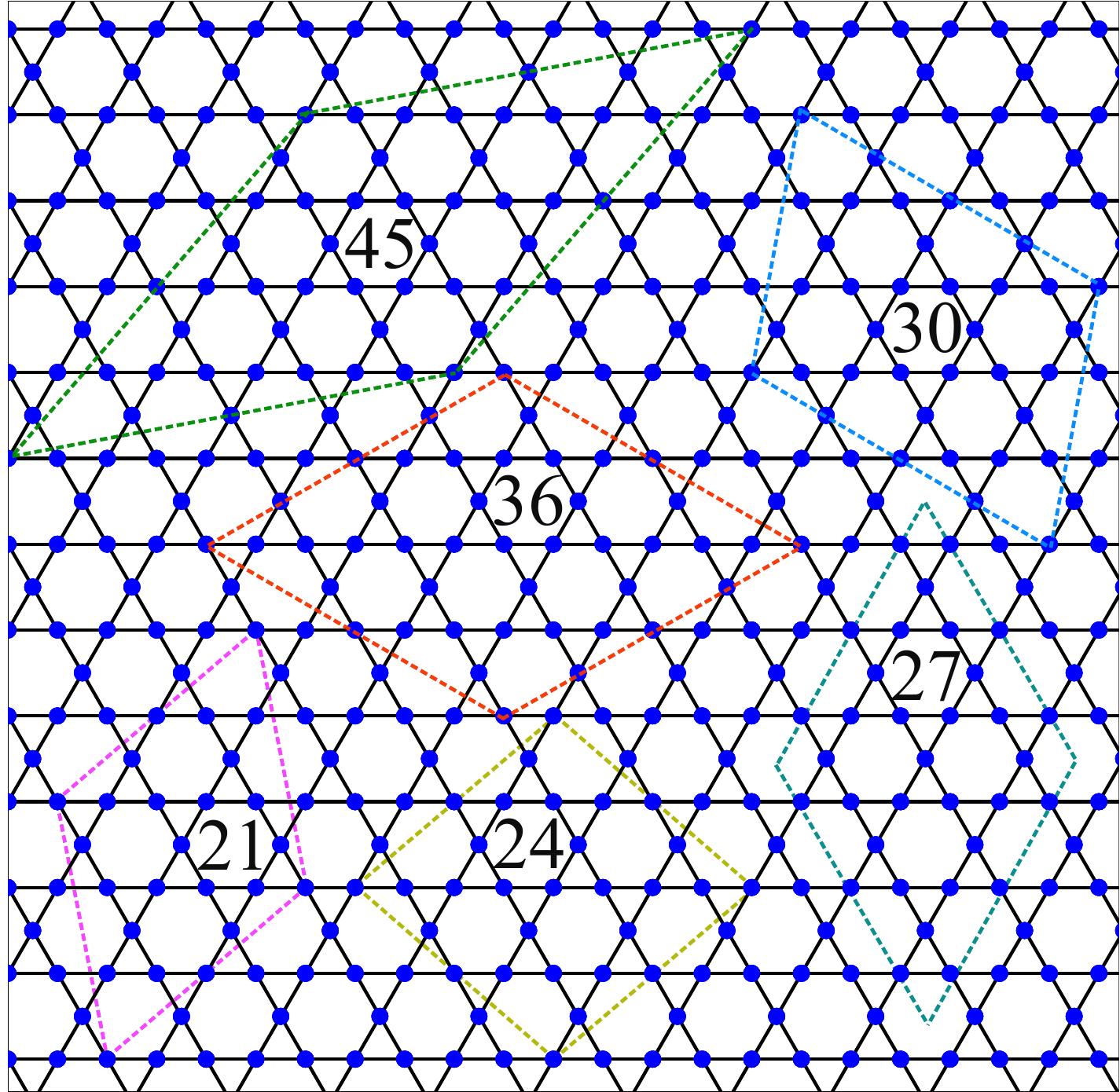}
\caption{
Cluster geometries used in the Lanczos and finite-temperature Lanczos calculations. Blue circles and black lines represent the sites and nearest-neighbor bonds of the kagome lattice, respectively. Colored dashed parallelograms indicate the finite clusters with PBCs, and the numbers denote the corresponding number of sites, $N=21, 24, 27, 30, 36,$ and $45$.
}
\label{S1}
\end{figure}

Figure~\ref{S1} shows the cluster geometries used in this study. The $N=21$, $24$, and $27$ clusters were used for calculations using the finite-temperature Lanczos method (FTLM), while the $N=24$, $27$, $30$, $36$, and $45$ clusters were used for zero-temperature Lanczos calculations. For the larger clusters, $N=30$, $36$, and $45$, the calculations were mainly performed in the high-magnetization region. The colored dashed parallelograms indicate the finite clusters with periodic boundary conditions (PBCs).

\vfill

\clearpage

\section{Finite-temperature Lanczos method}
In this study, finite-temperature properties were calculated using FTLM. 
We summarize the formulation used in FTLM calculations.

The Hamiltonian in a magnetic field is written as
\begin{equation}
  {\cal H}(h)={\cal H}_0-h\hat{M},
\end{equation}
where $\hat{M}=\sum_i S_i^z$ is the total magnetization. 
Since $\hat{M}$ is conserved, the Hilbert space is decomposed into fixed-magnetization sectors. 
We label each magnetization sector by the eigenvalue $M$ of $\hat{M}$, and denote the dimension of this sector by $N_{\rm st}(M)$.
The partition function is written as
\begin{equation}
  Z(T,h)=\sum_M e^{\beta hM} Z_M(T),
\end{equation}
where $\beta=1/T$ ($k_{\rm B}=1$) and
\begin{equation}
  Z_M(T)={\rm Tr}_M \exp(-\beta {\cal H}_0)
\end{equation}
is the sector-resolved partition function at zero field.

In the standard FTLM, the trace in each magnetization sector is evaluated by stochastic sampling with random vectors. For each normalized random vector $|v^{(r)}_M\rangle$ in the sector $M$, the Lanczos procedure is performed to construct an $N_{\rm L}$-dimensional Krylov subspace. Let $\epsilon_{j,M}^{(r)}$ and $|\psi_{j,M}^{(r)}\rangle$ be the Lanczos eigenvalues and eigenvectors obtained from the random vector $|v^{(r)}_M\rangle$. The sector-resolved partition function is approximated as
\begin{equation}
  Z_M(T) \simeq
  \frac{N_{\rm st}(M)}{R}
  \sum_{r=1}^{R}
  \sum_{j=0}^{N_{\rm L}-1}
  e^{-\beta \epsilon_{j,M}^{(r)}}
  \left|
  \langle v^{(r)}_M|\psi_{j,M}^{(r)}\rangle
  \right|^2 ,
\end{equation}
where $R$ is the number of random vectors.

The finite-temperature magnetization is obtained from
\begin{equation}
  M(T,h) =
  \frac{1}{Z(T,h)}
  \sum_M M e^{\beta hM} Z_M(T),
\end{equation}
and the normalized magnetization is $m(T,h)=M(T,h)/M_{\rm sat}$.

The zero-field magnetic susceptibility is calculated as
\begin{equation}
  \chi(T)  =
  \frac{1}{T}
  \frac{\sum_M M^2 Z_M(T)}
       {\sum_M Z_M(T)} .
\end{equation}

The internal energy at zero field is given by $E(T)=\langle {\cal H}_0\rangle$,
and the specific heat is calculated as
\begin{equation}
  C(T)=
  \frac{1}{T^2}
  \left[
  \langle {\cal H}_0^2\rangle
  -
  \langle {\cal H}_0\rangle^2
  \right].
\end{equation}

To improve the low-temperature accuracy, we also used the orthogonalized finite-temperature Lanczos method (OFTLM)~\cite{iFTLM}. In the standard FTLM, low-temperature quantities are sensitive to the accuracy of the lowest-energy contribution in the stochastic trace sampling. The OFTLM reduces this error by treating several low-lying eigenstates explicitly and performing the stochastic sampling only in the subspace orthogonal to them.

In each fixed-magnetization sector, we first calculate the lowest $N_{\rm V}$ eigenstates of ${\cal H}_0$,
\begin{equation}
  {\cal H}_0 |\Psi_{\nu,M}\rangle
  =
  E_{\nu,M} |\Psi_{\nu,M}\rangle .
\end{equation}
Here, $M$ labels the fixed-magnetization sector, and $\nu=0,\ldots,N_{\rm V}-1$. The projection operator onto the subspace spanned by these low-lying states is defined as
\begin{equation}
  P_M =
  \sum_{\nu=0}^{N_{\rm V}-1}
  |\Psi_{\nu,M}\rangle \langle \Psi_{\nu,M}| .
\end{equation}
The projection operator onto the orthogonal subspace is
\begin{equation}
  Q_M = 1-P_M .
\end{equation}
For each normalized random vector $|v^{(r)}_M\rangle$, we construct
\begin{equation}
  |v^{(r)\perp}_M\rangle =
  \frac{Q_M |v^{(r)}_M\rangle}
       {\sqrt{\langle v^{(r)}_M|Q_M|v^{(r)}_M\rangle}} .
\end{equation}
The Lanczos procedure is then performed starting from $|v^{(r)\perp}_M\rangle$.

The sector-resolved partition function is estimated in OFTLM as the sum of the exact contribution from the explicitly calculated low-lying states and the stochastic contribution from the remaining orthogonal subspace:
\begin{equation}
  Z_M^{\mathrm{OFTLM}}(T) =
  Z_M^{\rm ex}(T)+Z_M^{\perp}(T) .
\end{equation}
The first term is given by
\begin{equation}
  Z_M^{\rm ex}(T)
  =
  \sum_{\nu=0}^{N_{\rm V}-1}
  e^{-\beta E_{\nu,M}} .
\end{equation}
The second term is approximated by the Lanczos sampling in the orthogonal subspace,
\begin{equation}
  Z_M^{\perp}(T)
  \simeq
  \frac{N_{\rm st}(M)-N_{\rm V}}{R}
  \sum_{r=1}^{R}
  \sum_{j=0}^{N_{\rm L}-1}
  e^{-\beta \tilde{\epsilon}_{j,M}^{(r)}}
  \left|
  \langle v^{(r)\perp}_M|
  \tilde{\psi}_{j,M}^{(r)}\rangle
  \right|^2 .
\end{equation}
Here, $R$ is the number of random vectors, $N_{\rm L}$ is the Krylov dimension, and $\tilde{\epsilon}_{j,M}^{(r)}$ and $|\tilde{\psi}_{j,M}^{(r)}\rangle$ are the Lanczos eigenvalues and eigenvectors obtained from $|v^{(r)\perp}_M\rangle$.

In the present calculations, OFTLM was used for the $N=21$ and $24$ clusters to improve the accuracy of low-temperature thermodynamic quantities.
For the $N=24$ cluster, we used $N_V=3$, $N_L=140$, and $R=6$ for the magnetization sectors $M=0,\ldots,19$. For the high-magnetization sectors $20\le M\le24$, full exact diagonalization was performed. 
For the $N=21$ cluster, we used $N_V=4$, $N_L=100$, and $R=20$ for the magnetization sectors $M=0,\ldots,15$. For the high-magnetization sectors $16\le M\le21$, full exact diagonalization was performed.

For the $N=27$ cluster, we used the replaced finite-temperature Lanczos method (RFTLM)~\cite{iFTLM} only in the $M=0$ sector. The purpose of this treatment is to improve the low-temperature accuracy by replacing the stochastic FTLM contribution from the exactly known low-lying states with their exact contribution. In the present case, the $M=0$ ground state of the $N=27$ cluster is exactly twofold degenerate. Therefore, only this twofold-degenerate ground-state contribution was replaced.

 In the standard FTLM, the sector-resolved partition function at $M=0$ is approximated as
\begin{equation}
\begin{aligned}
Z_0^{\mathrm{FTLM}}(T)
&=
\frac{N_{\rm st}(0)}{R}
\sum_{r=1}^{R}
\sum_{j=0}^{N_{\mathrm{L}}-1}
e^{-\beta \epsilon_{j,0}^{(r)}}
\left|
\langle v^{(r)}_{0}|\psi_{j,0}^{(r)}\rangle
\right|^2 .
\end{aligned}
\end{equation}

Let $E_{0,0}$ be the ground-state energy in the $M=0$ sector, and let $|\Phi_{1,0}\rangle$ and $|\Phi_{2,0}\rangle$ be the two exactly degenerate ground states. The stochastic FTLM contribution from these two ground states is
\begin{equation}
\begin{aligned}
Z_{0,\mathrm{gs}}^{\mathrm{FTLM}}(T)
&=
\frac{N_{\rm st}(0)}{R}
\sum_{r=1}^{R}
\sum_{\alpha=1}^{2}
e^{-\beta E_{0,0}}
\left|
\langle v^{(r)}_{0}|\Phi_{\alpha,0}\rangle
\right|^2 .
\end{aligned}
\end{equation}
In RFTLM, this stochastic contribution is replaced by the exact twofold-degenerate ground-state contribution,
\begin{equation}
\begin{aligned}
Z_{0,\mathrm{gs}}^{\mathrm{exact}}(T)
&=
2 e^{-\beta E_{0,0}} .
\end{aligned}
\end{equation}
Thus, the RFTLM estimate of the partition function in the $M=0$ sector is given by
\begin{equation}
\begin{aligned}
Z_0^{\mathrm{RFTLM}}(T)
&=
Z_0^{\mathrm{FTLM}}(T)
-
Z_{0,\mathrm{gs}}^{\mathrm{FTLM}}(T)
+
Z_{0,\mathrm{gs}}^{\mathrm{exact}}(T) .
\end{aligned}
\end{equation}
This replacement was applied only to the twofold-degenerate ground-state contribution in the $M=0$ sector of the $N=27$ cluster. For the $N=27$ cluster, we used $R=20$ and $N_L=300$. For the sectors $M=1,\ldots,22$, standard FTLM was used. For the high-magnetization sectors $23\le M\le27$, full exact diagonalization was performed.

\section{Gaussian-kernel smoothing method}

In the Gaussian-kernel smoothing method, a smooth energy-density function is reconstructed from the discrete energy densities $e(m_i)$ obtained by Lanczos calculations in fixed-magnetization sectors~\cite{GK}. As described in the main text, we approximate the energy density as
\begin{equation}
  e_{\rm ker}(m) = b_0 + b_1 m + b_2 m^2 + b_3 m^3
  + \sum_i a_i \left[ \exp\left[-\frac{(m-m_i)^2}{2\ell^2}\right] + \exp\left[-\frac{(m+m_i)^2}{2\ell^2}\right] \right].
  \label{kernel}
\end{equation}
Here, $m_i$ denotes the discrete magnetization values obtained from the Lanczos calculations, $a_i$ are the coefficients of the Gaussian kernels, and $b_0,b_1,b_2,$ and $b_3$ are the coefficients of the polynomial part. The parameter $\ell$ is a hyperparameter that controls the width of the Gaussian kernels. The second Gaussian term is introduced to include the mirror contribution associated with $m\to -m$.

The coefficients are determined by solving a set of linear equations so that the energy data $y_i=e(m_i)$ are reproduced. We first define the Gaussian-kernel matrix as
\begin{equation}
K_{ij} = \exp\left\{ -\frac{(m_i-m_j)^2}{2\ell^2} \right\} + \exp\left\{ -\frac{(m_i+m_j)^2}{2\ell^2} \right\}, \qquad i,j=1,\ldots,n . 
\end{equation}
The matrix corresponding to the polynomial part is given by
\begin{equation}
P=
\begin{pmatrix}
1 & m_1 & m_1^2 & m_1^3 \\
1 & m_2 & m_2^2 & m_2^3 \\
\vdots & \vdots & \vdots & \vdots \\
1 & m_n & m_n^2 & m_n^3
\end{pmatrix}.
\end{equation}
By writing the unknown coefficients as $a=(a_1,\ldots,a_n)^T$ and $b=(b_0,b_1,b_2,b_3)^T$, and the energy data as $y=(y_1,\ldots,y_n)^T$, the coefficients without additional physical constraints are obtained by solving

\begin{equation}
\begin{pmatrix} 
K+\lambda I_n & P \\ P^T & 0 \end{pmatrix} 
\begin{pmatrix} a \\ b \end{pmatrix} = \begin{pmatrix} y \\ 0
\end{pmatrix}. 
\end{equation}
Here, $\lambda$ is a regularization parameter introduced to suppress excessively large Gaussian-kernel coefficients and to improve the numerical stability of the interpolation.

For the spin-1 kagome-lattice antiferromagnet, the Gaussian-kernel smoothing was performed using the lowest energies in fixed-magnetization sectors obtained for the $N=27, 30, 36,$ and $45$ clusters. 
Since the Lanczos results indicate plateaus at $m=0$, $1/3$, $7/9$, and $8/9$, these magnetization values were treated as candidate plateau positions in the smoothing procedure. We therefore did not approximate the entire magnetization range by a single smooth function. 
Instead, the magnetization range was divided into three regions,
\begin{equation}
0\le m<\frac{1}{3}, \qquad \frac{1}{3}\le m\le\frac{7}{9}, \qquad \frac{7}{9}<m\le\frac{8}{9},
\end{equation}
and the Gaussian-kernel smoothing was independently applied to each region. In the high-magnetization region $8/9\le m\le 1$, the energy is exactly determined from the localized-magnon state. Therefore, instead of applying numerical smoothing, we used the exact expression
\begin{equation}
e(m)=6m-4.
\end{equation}
The positions and presence of magnetization jumps were not imposed in advance; they were obtained by minimizing $e_{\rm ker}(m)-hm$ with respect to $m$ at each magnetic field.

For comparison, we performed benchmark calculations for the spin-1/2 one-dimensional Heisenberg chain with an $N=24$ periodic cluster and for the spin-1/2 triangular-lattice Heisenberg antiferromagnet with an $N=36$ periodic cluster. 
Since no magnetization plateau is expected at $m=0$ in these systems, we set $b_1=0$.
In these benchmark calculations, we further imposed two physical constraints at the saturation magnetization. The first constraint fixes the energy density at $m=1$ to the exact energy density of the fully polarized state, $e_{\rm sat}$. 
The second constraint fixes the slope of the energy curve at $m=1$ to the saturation field $h_{\rm s}$. 
Namely, we imposed 
\begin{equation} 
e_{\rm ker}(1)=e_{\rm sat}, \qquad \left. \frac{d e_{\rm ker}(m)}{dm} \right|_{m=1} = h_{\rm s}. 
\end{equation} 
These two constraints were incorporated as additional linear conditions when solving for the kernel and polynomial coefficients. For the triangular-lattice case, since a magnetization plateau exists at $m=1/3$, the magnetization range was divided into two intervals, 
\begin{equation} 
0\le m\le \frac{1}{3}, \qquad \frac{1}{3}\le m\le 1, 
\end{equation} 
and the Gaussian-kernel smoothing was independently performed in each interval.

\begin{figure}[tb]
\centering
\includegraphics[width=156mm]{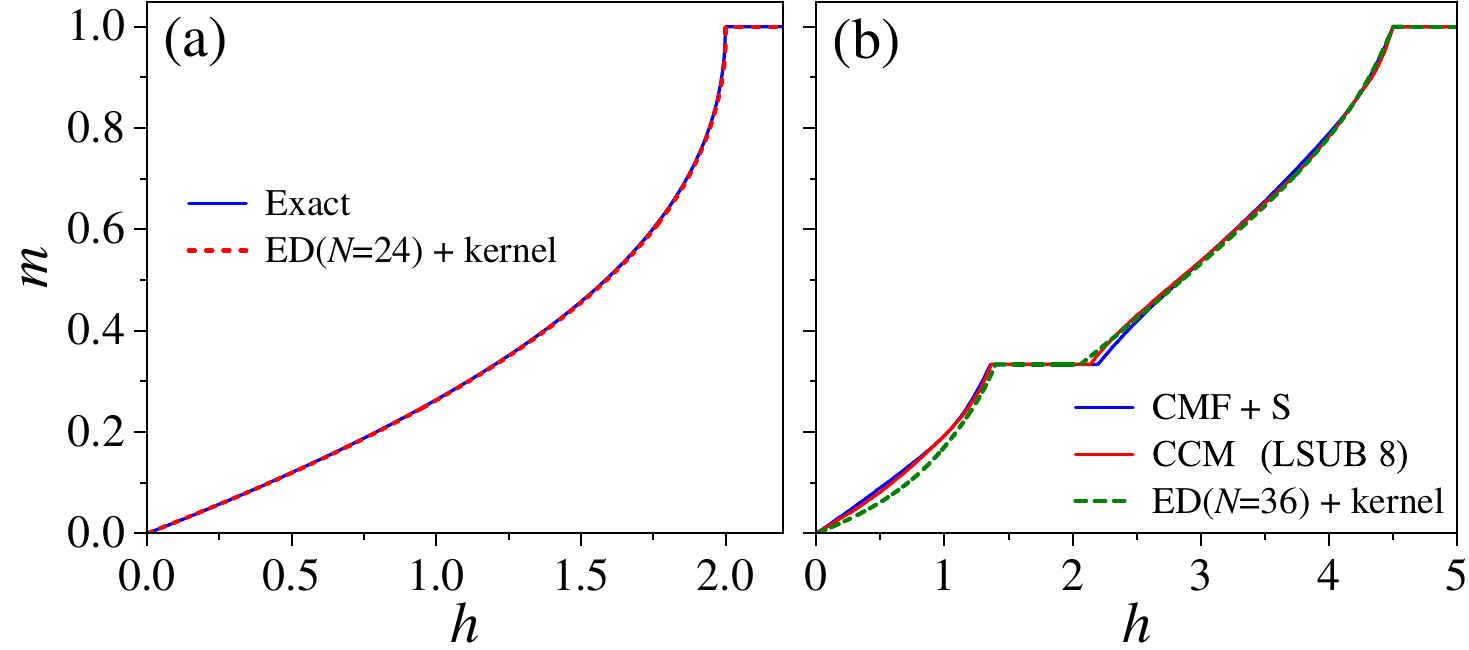}
\caption{
Benchmark magnetization curves obtained by the Gaussian-kernel smoothing method. (a) Spin-1/2 one-dimensional Heisenberg chain. 
The blue solid line denotes the exact result, and the red dashed line denotes the result obtained from the $N=24$  cluster using the Gaussian-kernel smoothing method. 
(b) Spin-1/2 triangular-lattice Heisenberg antiferromagnet. 
The blue solid and red solid lines denote the results obtained by CMF+S and CCM-LSUB8, respectively. The green dashed line denotes the result obtained from the $N=36$ cluster using the Gaussian-kernel smoothing method.
}
\label{S2}
\end{figure}

Figure~\ref{S2} shows benchmark results for the Gaussian-kernel smoothing method. For the spin-1/2 one-dimensional Heisenberg chain, as shown in Fig.~\ref{S2} (a), the magnetization curve obtained by the present method agrees very well with the exact result. For the spin-1/2 triangular-lattice Heisenberg antiferromagnet, as shown in Fig.~\ref{S2} (b), the obtained magnetization curve also shows good agreement with highly accurate reference results obtained by the cluster mean-field method with scaling (CMF+S)~\cite{CMF} and the coupled cluster method within the LSUB8 approximation (CCM-LSUB8)~\cite{CCM}. These results demonstrate that the present method is effective for reconstructing smooth magnetization curves from discrete finite-size data. The hyperparameters used here are $\ell=0.4$ and $\lambda=1.0\times10^{-6}$ for the one-dimensional chain, and $\ell=0.4$ and $\lambda=2.0\times10^{-5}$ for the triangular lattice.

The result of the Gaussian-kernel smoothing method for the spin-1 kagome-lattice Heisenberg antiferromagnet is shown in Fig.~1 of the main text. In this calculation, we used the hyperparameters $\ell=0.3$ and $\lambda=1.0\times10^{-6}$. In Fig.~7(a) of the main text, the zero-temperature magnetization curve obtained by the present method is compared with the low-temperature magnetization curves obtained by FTLM at $T=0.05$ and $T=0.1$. The zero-temperature curve obtained by the present method is consistent with the low-temperature FTLM results, although the sharp features in the zero-temperature curve are naturally rounded at finite temperatures. This agreement indicates that a smooth zero-temperature magnetization curve is appropriately reconstructed from the discrete energy data in fixed-magnetization sectors. The hyperparameters were chosen by also taking into account consistency with the low-temperature FTLM results.

\section{Finite-size dependence of finite-temperature magnetization curves}

\begin{figure}[tb]
\centering
\includegraphics[width=156mm]{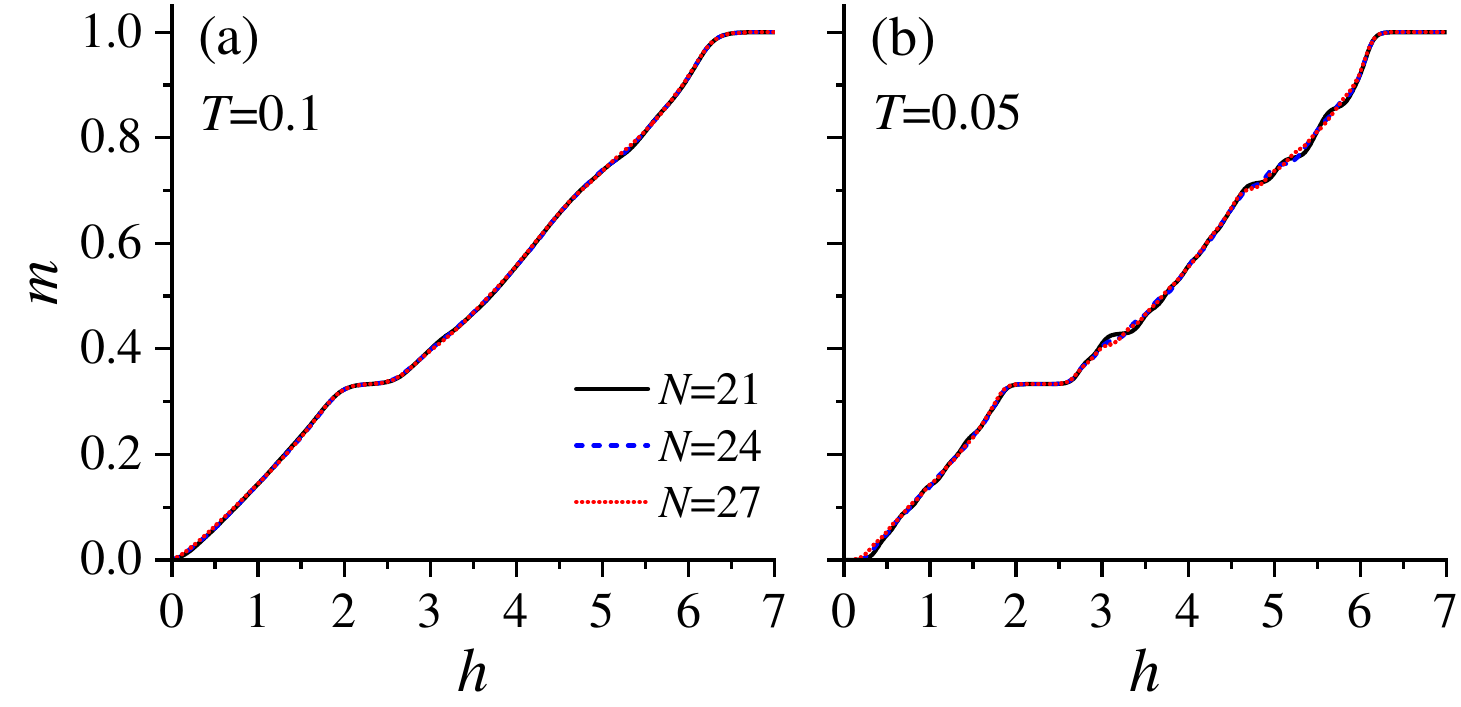}
\caption{
Finite-size dependence of the finite-temperature magnetization curves obtained using FTLM.  
The results are shown for the $N=21$, 24, and 27 clusters at (a) $T=0.1$ and (b) $T=0.05$.
}
\label{S3}
\end{figure}

Figure~\ref{S3} shows the finite-size dependence of the finite-temperature magnetization curves. In Fig.~\ref{S3} (a), we compare the results for the $N=21$, 24, and 27 clusters at $T=0.1$. The three magnetization curves almost completely overlap, indicating that finite-size effects are very small at this temperature. Thus, for $T\simeq 0.1$, the FTLM provides magnetization curves that are already close to the thermodynamic-limit behavior.
In contrast, the results at $T=0.05$, shown in Fig.~\ref{S3} (b), exhibit visible differences among the cluster sizes. This indicates that the discrete energy-level structure of finite-size systems remains relevant at lower temperatures and affects the magnetization curves. These results suggest that, for the cluster sizes used in this study, the finite-temperature magnetization curves are reliable for $T\ge 0.1$, whereas finite-size effects should be treated with care in the lower-temperature region $T<0.1$.

\clearpage

\end{document}